\documentclass{revtex4}
\usepackage{amssymb}
\usepackage{amsfonts}
\usepackage{amsmath}

\input{tcilatex}

\begin{document}

\title{Position-dependent mass momentum operator and minimal coupling: point
canonical transformation and isospectrality}
\author{Omar Mustafa }
\email{omar.mustafa@emu.edu.tr}
\author{Zeinab Algadhi}
\email{zeinab.algadhi@emu.edu.tr}
\affiliation{Department of Physics, Eastern Mediterranean University, G. Magusa, north
Cyprus, Mersin 10 - Turkey,\\
Tel.: +90 392 6301378; fax: +90 3692 365 1604.}

\begin{abstract}
\textbf{Abstract:} The classical and quantum mechanical correspondence for
constant mass settings is used, along with some point canonical
transformation, to find the position-dependent mass (PDM) classical and
quantum Hamiltonians. The comparison between the resulting quantum
PDM-Hamiltonian and the von Roos PDM-Hamiltonian implied that the ordering
ambiguity parameters of von Roos are strictly determined. Eliminating, in
effect, the ordering ambiguity associated with\ the von Roos
PDM-Hamiltonian. This, consequently, played a vital role in the construction
and identification of the PDM-momentum operator. The same recipe is followed
to identify the form of the minimal coupling of electromagnetic interactions
for the classical and quantum PDM-Hamiltonians. It turned out that whilst
the minimal coupling may very well inherit the usual form in classical
mechanics (i.e., $p_{_{j}}\left( \vec{x}\right) \longrightarrow
p_{_{j}}\left( \vec{x}\right) -e\,A_{j}\left( \vec{x}\right) $, where $%
p_{_{j}}\left( \vec{x}\right) $ is the $j$th component of the classical
PDM-canonical-momentum), it admits a\ necessarily different and vital form
in quantum mechanics (i.e., $\widehat{p}_{_{j}}\left( \vec{x}\right) /\sqrt{%
m\left( \vec{x}\right) }\longrightarrow \left( \widehat{p}_{_{j}}\left( \vec{%
x}\right) -e\,A_{j}\left( \vec{x}\right) \right) /\sqrt{m\left( \vec{x}%
\right) }$, where $\widehat{p}_{_{j}}\left( \vec{x}\right) $ is the $j$th
component of the quantum PDM-momentum operator). Under our point
transformation settings, only one of the two commonly used vector potentials
(i.e., $\overrightarrow{A}\left( \vec{x}\right) \sim (-x_{_{2}},x_{_{1}},0)$%
) is found eligible and is considered for our Illustrative examples.


\end{abstract}

\maketitle

\section{Introduction}

When the information on the material properties is encoded in the mass of a
quantum particles, the concept of quantum particles endowed with
position-dependent mass (PDM) becomes unavoidable. To deal with such quantum
mechanical problems, von Roos \cite{1} has suggested a PDM Hamiltonian of
the form 
\begin{equation}
\hat{H}=-\frac{1}{4}\left[ M\left( \vec{x}\right) ^{\alpha }\partial
_{x_{j}}M\left( \vec{x}\right) ^{\beta }\partial _{x_{j}}M\left( \vec{x}%
\right) ^{\gamma }+M\left( \vec{x}\right) ^{\gamma }\partial _{x_{j}}M\left( 
\vec{x}\right) ^{\beta }\partial _{x_{j}}M\left( \vec{x}\right) ^{\alpha }%
\right] +V\left( \vec{x}\right) .
\end{equation}%
Where $M\left( \vec{x}\right) =m_{\circ }m\left( \vec{x}\right) $, $m_{\circ
}$ is the rest mass, , $m\left( \vec{x}\right) $ is a position-dependent
dimensionless scalar multiplier that forms the position-dependent mass $%
M\left( \vec{x}\right) $, $\vec{x}=\left( x_{_{1}},x_{_{2}},x_{_{3}}\right) $%
, $\partial _{x_{j}}=\partial /\partial x_{j}$, $j=1,2,3$, $V\left( \vec{x}%
\right) $ is the potential force field, and the summation runs over repeated
indices, unless otherwise mentioned. This Hamiltonian has been a subject of
interest over the last few decades not only as a mathematically challenging
Hamiltonian but also as a feasibly applicable one in many fields of physics.
Obviously, a parametric ordering ambiguity (i.e., in $\alpha ,\beta ,$ and $%
\gamma $) arises in the formation of such Hamiltonian as a consequence of
the non-unique representation of the kinetic energy term. It is known that,
the ordering ambiguity parameters $\alpha ,\beta ,$ and $\gamma $ only
satisfy the von Roos constraint $\alpha +\beta +\gamma =-1$. A constraint
that is, in fact, manifested by the requirement that the von Roos
Hamiltonian should collapse into the constant mass Hamiltonian settings as $%
m\left( \vec{x}\right) =1$.

Many attempts were made to come out with a physically acceptable parametric
ordering settings \cite{2,3,4,5,6,7,8,9,10,11,12}. The only physically
acceptable condition (along with the von Roos constraint) on the ambiguity
parameters is that $\alpha =\gamma $ to ensure the continuity condition at
the abrupt heterojunction between two crystals (c.f., e.g., Ref. \cite{13}).
The rest were either based on circumstantial ordering settings that fit into
exact solvability requirement, or those of mathematical and classical
mechanical challenging nature \cite%
{14,15,16,17,18,19,20,21,22,23,24,25,26,27,28,29,30,31,32,33,34,35,36,37,38,39,40,41,42,43,44,45,46,47,48}%
, or even those that are based on an intelligent guess of the factorization
structure of the kinetic energy operator (c.f., e.g., \cite{49,50} and
related references cited therein). To the best of our knowledge, however, no
attempts were ever made to construct and identify the PDM-momentum operator.
We do this as part of the current methodical proposal. Nevertheless, it
should be noted that Hamiltonian (1) would, in a straightforward manner,
imply a time-independent PDM Schr\"{o}dinger equation (in $\hbar =2m_{\circ
}=c=1$ units) of the form%
\begin{gather}
\left\{ -\frac{1}{m\left( \vec{x}\right) }\,\partial _{x_{j}}^{2}+\left[ 
\frac{\partial _{x_{j}}m\left( \vec{x}\right) }{m\left( \vec{x}\right) ^{2}}%
\right] \,\partial _{x_{j}}-\left[ \alpha \left( \alpha +\beta +1\right)
+\beta +1\right] \left( \frac{\left[ \partial _{x_{j}}m\left( \vec{x}\right) %
\right] ^{2}}{m\left( \vec{x}\right) ^{3}}\right) \right.  \notag \\
\left. +\frac{1}{2}\left( 1+\beta \right) \left[ \frac{\partial
_{x_{j}}^{2}m\left( \vec{x}\right) }{m\left( \vec{x}\right) ^{2}}\right]
+V\left( \vec{x}\right) \right\} \phi \left( \vec{x}\right) =E\phi \left( 
\vec{x}\right) .
\end{gather}%
This equation is to play a critical role in the determination of the
ambiguity parameters and consequently in the construction of the
PDM-momentum operator as well as in the identification of the minimal
coupling of electromagnetic interactions. The organization of the current
methodical proposal is in the respective order, therefore.

In section II, we start with the Lagrangian of a classical particle of mass $%
m_{\circ }$ moving in a scalar potential field $V\left( \vec{q}\right) $, in
the generalized coordinates $\vec{q}=\left(
q_{_{1}},q_{_{2}},q_{_{3}}\right) $, to build up the classical and
consequently the quantum mechanical Hamiltonians. Based on our very recent
work on the so called point canonical transformation \cite{39}, we detail
out the mapping(s)/connection(s) between the quantum mechanical PDM-Schr\"{o}%
dinger equation (2) and the apparently standard textbook Schr\"{o}dinger
equation for constant mass $m_{\circ }$ in\ the generalized coordinates.Once
the mapping is made clear, the ordering ambiguity in (2) disappears and the
parametric setting become strictly determined. However, a question of
delicate nature arises in the process as to "\textit{what is the form of the
position-dependent mass momentum operator, if there is any at all?". }In
section III, we find that the answer to this question in the very
fundamentals of "\textit{Quantum Mechanics" }of \ S. Gasirowicz \cite{51}
(equation (17) and (18) below). In so doing, we first find the so called PDM
pseudo-momentum operator $\widehat{\pi }_{_{j}}\left( \vec{q}\left( \vec{x}%
\right) \right) $ and connect it with the PDM-momentum operator through $%
\widehat{P}_{_{j}}\left( \vec{x}\right) =\sqrt{m\left( \vec{x}\right) }%
\widehat{\pi }_{_{j}}\left( \vec{q}\left( \vec{x}\right) \right) $.
Surprisingly, it turns out that the construction of the PDM pseudo-momentum
operator (hence, PDM-momentum operator) has nothings to do with the
ambiguity parameters but it only depends on the transformation of the wave
function form the generalized coordinate $\vec{q}=\left(
q_{_{1}},q_{_{2}},q_{_{3}}\right) $ into the rectangular coordinates $\vec{x}%
=\left( x_{_{1}},x_{_{2}},x_{_{3}}\right) $ (i.e., $\psi \left( \vec{q}%
\right) \rightarrow m\left( \vec{x}\right) ^{\upsilon }\phi \left( \vec{x}%
\right) ,$ equation (14) below.

Next, having the PDM-momentum (operator) identified in both the classical
and quantum mechanical forms, we dwell on the nature of the minimal coupling
of electromagnetic interactions for PDM-settings in section IV. Therein, we
find that the simplest way of coupling the electromagnetic interaction is to
take the Hamilton's PDM pseudo-momentum $\pi _{_{j}}\left( \vec{q}\left( 
\vec{x}\right) \right) $ as the sum of the PDM pseudo-kinetic momentum $%
m_{\circ }\tilde{q}_{_{j}\,}=m_{\circ }\left( \sqrt{m\left( \vec{x}\right) }%
\dot{x}_{_{j}}\right) $ and $e\,A_{j}\left( \vec{q}\left( \vec{x}\right)
\right) $ ( i.e., $\pi _{_{j}}\left( \vec{q}\left( \vec{x}\right) \right)
=m_{\circ }\left( \sqrt{m\left( \vec{x}\right) }\dot{x}_{_{j}}\right)
+e\,A_{j}\left( \vec{q}\left( \vec{x}\right) \right) $). Hence, the minimal
coupling of the electromagnetic interactions turns out to be $\pi
_{_{j}}\left( \vec{q}\left( \vec{x}\right) \right) \longrightarrow \pi
_{_{j}}\left( \vec{q}\left( \vec{x}\right) \right) -e\,A_{j}\left( \vec{q}%
\left( \vec{x}\right) \right) $ and \ \ $E\longrightarrow E-e\,\varphi (\vec{%
q}\left( \vec{x}\right) )$ for the Classical PDM-Hamiltonian. Which, in
terms of the canonical PDM-momentum, reads $p_{_{_{j}}}\left( \vec{x}\right)
/\sqrt{m\left( \vec{x}\right) }\longrightarrow \left( p_{_{_{j}}}\left( \vec{%
x}\right) -e\,A_{j}\left( \vec{x}\right) \right) /\sqrt{m\left( \vec{x}%
\right) }$. Consequently, for the quantum mechanical PDM-Hamiltonian $%
\widehat{p}_{_{_{j}}}\left( \vec{x}\right) /\sqrt{m\left( \vec{x}\right) }%
\longrightarrow \left( \widehat{p}_{_{_{j}}}\left( \vec{x}\right)
-e\,A_{j}\left( \vec{x}\right) \right) /\sqrt{m\left( \vec{x}\right) }$ and $%
E\longrightarrow E-e\,\varphi (\vec{q}\left( \vec{x}\right) )$. Obviously,
the minimal coupling for the classical PDM-Hamiltonian effectively remains
in the same form as that for constant mass setting (i.e., $p_{_{_{j}}}\left( 
\vec{x}\right) \longrightarrow p_{_{_{j}}}\left( \vec{x}\right)
-e\,A_{j}\left( \vec{x}\right) $, for it does not indulge any differential
operator), whereas for the quantum PDM-Hamiltonian the kinetic energy
operator should be correctly presented as%
\begin{equation*}
\widehat{T}=\left( \frac{\widehat{p}_{_{_{j}}}\left( \vec{x}\right) }{\sqrt{%
m\left( \vec{x}\right) }}\right) ^{2}\Longrightarrow \widehat{T}=\left( 
\frac{\widehat{p}_{_{_{j}}}\left( \vec{x}\right) -e\,A_{j}\left( \vec{x}%
\right) }{\sqrt{m\left( \vec{x}\right) }}\right) ^{2}\neq \frac{\left( 
\widehat{p}_{_{_{j}}}\left( \vec{x}\right) -e\,A_{j}\left( \vec{x}\right)
\right) ^{2}}{m\left( \vec{x}\right) }
\end{equation*}%
where $\widehat{p}_{_{_{j}}}\left( \vec{x}\right) $ is the $j$th component
of the PDM-momentum operator, $E$ is the eigenenergy and $e\,\varphi (\vec{x}%
)$ is the scalar part of the electromagnetic four vector potential $A_{\mu
}= $ $\left( \vec{A}(\vec{x}),i\varphi (\vec{x})\right) $. In section V we
test the eligibility of the commonly used vector potentials and single out $%
\overrightarrow{A}\left( \vec{x}\right) \sim (-x_{_{2}},x_{_{1}},0)$ as the
only eligible vector potential within our current methodical proposal
settings, of course. In section VI we give two illustrative examples that
include magnetic and electric fields. We conclude in section VII.

\section{Point transformation and Classical-Quantum correspondence}

Consider the motion of a classical particle of a constant rest mass $%
m_{\circ }$ moving in a potential field $V(\vec{q})$, where $\vec{q}=\left(
q_{_{1}},q_{_{2}},q_{_{3}}\right) =q_{_{1}}\widehat{q}_{_{1}}+q_{_{2}}%
\widehat{q}_{_{2}}+q_{_{3}}\widehat{q}_{_{3}}$ are the generalized
coordinates. The corresponding Lagrangian for such a system is given by%
\begin{equation}
L\left( q_{_{j}},\tilde{q}_{_{j}};\tau \right) =\frac{1}{2}m_{\circ }\tilde{q%
}_{_{j}}^{2}-V(\vec{q})\text{ };\text{ \ }\tilde{q}_{_{j}}=\frac{dq_{_{j}}}{%
d\tau };\text{ }\,j=1,2,3.
\end{equation}%
Where $\tau $ is a re-scaled time \cite{39} and $L\left( q_{_{j}},\tilde{q}%
_{_{j}};\tau \right) =L\left( q_{_{1}},q_{_{2}},q_{_{3}},\tilde{q}_{_{1}},%
\tilde{q}_{_{2}},\tilde{q}_{_{3}};\tau \right) $ is to be used for the
economy of notations. Under such settings, the classical Hamiltonian reads%
\begin{equation}
H\left( q_{_{j}},P_{j};\tau \right) =\tilde{q}_{_{j}\,}P_{j}-L\left(
q_{_{j}},\tilde{q}_{_{j}};\tau \right) =\frac{1}{2}m_{\circ }\tilde{q}%
_{_{j}}^{2}+V(\vec{q}),
\end{equation}%
and represents a constant of motion where $dH\left( q_{_{j}},P_{j};\tau
\right) /d\tau =0$ (c.f., e.g., Mustafa \cite{39}). Here, the $j$th
component of the canonical momentum (associated with the generalized
coordinate $q_{_{j}}$) 
\begin{equation}
P_{j}=\frac{\partial }{\partial \tilde{q}_{_{j}\,}}L\left( q_{_{j}},\tilde{q}%
_{_{j}};\tau \right) \Longrightarrow P_{j}=m_{\circ }\tilde{q}_{_{j}\,},
\end{equation}%
is used. However, the Hamiltonian is often realized to be a function of
position $q_{_{j}}$ and canonical momentum $P_{j}$ (and not a function of
position $q_{_{j}}$ and velocity $\tilde{q}_{_{j}\,}$). It is more
appropriate, therefore, to re-cast the classical Hamiltonian (4) as%
\begin{equation}
H\left( q_{_{j}},P_{j};\tau \right) =\frac{P_{j}^{2}}{2m_{\circ }}+V(\vec{q}%
).
\end{equation}%
Hence, the corresponding quantum mechanical Hamiltonian is obtained by the
identification of the $j$th canonical momentum $P_{j}$ with the operator $%
\widehat{P}_{j}=-i\partial /\partial q_{_{j}}=-i\partial _{q_{_{j}}}$, that
satisfies the canonical commutation relations $\left[ q_{_{i}},\widehat{P}%
_{j}\right] =-i\left[ q_{_{i}},\partial _{q_{_{j}}}\right] =i\delta _{ij}$
and consequently yields ( with $\hbar =2m_{\circ }=c=1$) to%
\begin{equation}
\widehat{H}\left( q_{_{j}},P_{j};\tau \right) =-\partial _{q_{_{j}}}^{2}+V(%
\vec{q}).
\end{equation}%
Then, the corresponding time-independent Schr\"{o}dinger equation reads%
\begin{equation}
\left\{ -\partial _{q_{_{j}}}^{2}+V(\vec{q})\right\} \psi \left( \vec{q}%
\right) =\Lambda \psi \left( \vec{q}\right)
\end{equation}

At this very point, we would like to figure out the mapping(s)/connection(s)
between the quantum mechanical PDM-Schr\"{o}dinger equation (2) and the
apparently standard textbook Schr\"{o}dinger equation for constant mass in
(8) (equation (8) is obtained as a quantum mechanical correspondence of the
classical Hamiltonian (6)). To do that, we invest in our experience on the
point transformation very recently suggested by Mustafa \cite{39} and define 
\begin{equation}
dq_{_{i}}=\delta _{ij}\sqrt{g\left( \vec{x}\right) }\,dx_{_{j}}=\sqrt{%
g\left( \vec{x}\right) }\,dx_{_{i}}\Longrightarrow \frac{\partial q_{_{i}}}{%
\partial x_{_{j}}}=\delta _{ij}\sqrt{g\left( \vec{x}\right) }\Longrightarrow
q_{_{j}}=\int \sqrt{g\left( \vec{x}\right) }\,dx_{_{j}},\text{ \ }\tau
=\dint f\left( \vec{x}\right) dt.
\end{equation}%
No summation over repeated index holds in (9). Therefore, this type of
transformation necessarily means that the differential change in $q_{_{j}}$\
is defined through the matrix%
\begin{equation}
\left[ 
\begin{tabular}{l}
$dq_{_{1}}$ \\ 
$dq_{_{2}}$ \\ 
$dq_{_{3}}$%
\end{tabular}%
\right] =\left[ 
\begin{tabular}{ccc}
$\frac{\partial q_{_{1}}}{\partial x_{1}}$ & 0 & 0 \\ 
0 & $\frac{\partial q_{_{2}}}{\partial x_{2}}$ & 0 \\ 
0 & 0 & $\frac{\partial q_{_{3}}}{\partial x_{_{3}}}$%
\end{tabular}%
\right] \left[ 
\begin{tabular}{l}
$dx_{_{1}}$ \\ 
$dx_{_{2}}$ \\ 
$dx_{_{3}}$%
\end{tabular}%
\right] =\sqrt{g\left( \vec{x}\right) }\left[ 
\begin{tabular}{l}
$dx_{_{1}}$ \\ 
$dx_{_{2}}$ \\ 
$dx_{_{3}}$%
\end{tabular}%
\right] \Longrightarrow \text{\ }\tilde{q}_{_{_{j}}}=\frac{\sqrt{g\left( 
\vec{x}\right) }}{f\left( \vec{x}\right) }\dot{x}_{_{j}}\,;\text{ }\dot{x}%
_{_{j}}=\frac{dx_{_{j}}}{dt},
\end{equation}%
Consequently, the unit vectors in the direction of $q_{_{i}}$ are obtained
as 
\begin{equation}
\widehat{q}_{_{i}}=\sqrt{g\left( \vec{x}\right) }\left[ \left( \frac{%
\partial x_{_{1}}}{\partial q_{_{j}}}\right) \,\widehat{x}_{_{1}}+\left( 
\frac{\partial x_{_{2}}}{\partial q_{_{j}}}\right) \,\widehat{x}%
_{_{2}}+\left( \frac{\partial x_{_{3}}}{\partial q_{_{j}}}\right) \,\widehat{%
x}_{_{3}}\right] \Longrightarrow \widehat{q}_{_{i}}=\,\widehat{x}_{_{i}}.
\end{equation}%
Moreover, one should notice that such a point transformation recipe, along
with the condition $g\left( \vec{x}\right) =m\left( \vec{x}\right) f\left( 
\vec{x}\right) ^{2}$, would keep the related Euler-Lagrange equations
invariant (for more detailed analysis on this issue one may refer to Mustafa 
\cite{39}). Therefore, the related time-independent Schr\"{o}dinger equation
(8) would, with the substitutions of%
\begin{equation}
\psi \left( \vec{q}\right) =g\left( \vec{x}\right) ^{\upsilon }\phi \left( 
\vec{x}\right) ,
\end{equation}%
yield%
\begin{gather}
\left\{ -\frac{1}{g\left( \vec{x}\right) }\,\partial _{x_{j}}^{2}-\left(
2\upsilon -\frac{1}{2}\right) \left( \frac{\partial _{x_{j}}g\left( \vec{x}%
\right) }{g\left( \vec{x}\right) ^{2}}\right) \,\partial _{x_{j}}-\upsilon
\left( \upsilon -\frac{3}{2}\right) \left( \frac{\left[ \partial
_{x_{j}}g\left( \vec{x}\right) \right] ^{2}}{g\left( \vec{x}\right) ^{3}}%
\right) \right.  \notag \\
\left. -\upsilon \left( \frac{\partial _{x_{j}}^{2}g\left( \vec{x}\right) }{%
g\left( \vec{x}\right) ^{2}}\right) +V\left( \vec{q}\left( \vec{x}\right)
\right) \right\} \phi \left( \vec{x}\right) =\Lambda \phi \left( \vec{x}%
\right) .
\end{gather}

Nevertheless, we shall be interested in quantum mechanical systems in (8)
that are exactly solvable, conditionally exactly solvable, or quasi-exactly
solvable to reflect on the solvability of a given PDM system as that in
(13). Therefore, the eigenvalues $\Lambda $ of (13) should be not only
position-independent but also isospectral to $E$ of (2), i.e., $E=\Lambda $.
Under such settings, one immediately concludes that $f\left( \vec{x}\right)
=1\Longrightarrow $ $\tau =t$ and $g\left( \vec{x}\right) =m\left( \vec{x}%
\right) $ to keep the total energy position-independent and ensures
isospectrality between (2) and (8), Now, we compare the second term of (2)
with second term of (13) to imply that%
\begin{equation}
2\upsilon -\frac{1}{2}=-1\Longrightarrow \upsilon =-\frac{1}{4}%
\Longrightarrow \psi \left( \vec{q}\right) =m\left( \vec{x}\right)
^{-1/4}\phi \left( \vec{x}\right) .
\end{equation}%
Hence, equation (13) reduces to 
\begin{equation}
\left\{ -\frac{1}{m\left( \vec{x}\right) }\,\partial _{x_{j}}^{2}+\left[ 
\frac{\partial _{x_{j}}m\left( \vec{x}\right) }{m\left( \vec{x}\right) ^{2}}%
\right] \,\partial _{x_{j}}-\frac{7}{16}\left( \frac{\left[ \partial
_{x_{j}}m\left( \vec{x}\right) \right] ^{2}}{m\left( \vec{x}\right) ^{3}}%
\right) +\frac{1}{4}\left[ \frac{\partial _{x_{j}}^{2}m\left( \vec{x}\right) 
}{m\left( \vec{x}\right) ^{2}}\right] +V\left( \vec{q}\left( \vec{x}\right)
\right) \right\} \phi \left( \vec{x}\right) =E\phi \left( \vec{x}\right) ,
\end{equation}%
where $V\left( \vec{q}\left( \vec{x}\right) \right) =V\left( \vec{x}\right) $
of (2) ( which would, in the process, determine the form of $V\left( \vec{x}%
\right) $ for a given $V\left( \vec{q}\left( \vec{x}\right) \right) $ and
vice versa). Consequently, moreover, one obtains the identities 
\begin{equation}
\alpha \left( \alpha +\beta +1\right) +\beta +1=\frac{7}{16},\text{ \ }\frac{%
1}{2}\left( 1+\beta \right) =\frac{1}{4}.
\end{equation}%
Equations (8) and (15) are isospectral. Yet, the comparison clearly suggests
that the ordering ambiguity parameters are strictly determined in (16)
(along with the von Roos constraint $\alpha +\beta +\gamma =-1$) as $\beta
=-1/2$, and $\alpha =-1/4=\gamma $. The result that $\alpha =\gamma =-1/4$
satisfy the continuity condition at the abrupt heterojunction between two
crystals (c.f., e.g., Ref. \cite{13}). Hereby, we may safely conclude that
the PDM quantum mechanical correspondence of the PDM classical mechanical
settings removes the ordering ambiguity in the von Roos PDM-Hamiltonian (1).
We adopt this parametric result and proceed with the point transformation
settings used above.

\section{Construction of the PDM-momentum operator}

Having had the correlation between the Schr\"{o}dinger equation (in the
generalized coordinates) and the PDM-Schr\"{o}dinger equation (in the
rectangular coordinates) been identified, through (8)-(16), we now need to
address a question of delicate nature as to "\textit{what is the form of the
position-dependent mass momentum operator, if there is any at all?". }The
answer to this question very well lies in the very fundamentals of \textit{%
"Quantum mechanics"} by S. Gasiorowicz \cite{51}. Therein, the
one-dimensional quantum mechanical momentum operator $\widehat{p}%
_{x}=-i\partial /\partial x$ is determined through 
\begin{equation}
\left\langle p_{_{x}}\right\rangle =m_{\circ }\frac{d}{dt}\left\langle
x\right\rangle =\int\limits_{-\infty }^{\infty }dx\Psi ^{\ast }\left(
x,t\right) \left( -i\frac{\partial }{\partial x}\right) \Psi \left(
x,t\right) \Longrightarrow \widehat{p}_{_{x}}=-i\partial /\partial x.
\end{equation}%
This would suggest that the one-dimensional quantum momentum operator in the
generalized coordinate $q$ for the one-dimensional quantum mechanical system
is also obtainable through the same recipe as%
\begin{equation}
\left\langle P_{_{q}}\right\rangle =m_{\circ }\frac{d}{dt}\left\langle
q\right\rangle =\int\limits_{-\infty }^{\infty }dq\Psi ^{\ast }\left(
q,t\right) \left( -i\frac{\partial }{\partial q}\right) \Psi \left(
q,t\right) \Longrightarrow \widehat{P}_{_{q}}=-i\partial /\partial q.
\end{equation}%
Which is, in fact, what we have readily used above. Of course, the details
of the intermediate steps are straightforward and hold true for both (17)
and (18) and need not be reiterated here. Next, if we use the corresponding
one-dimensional point transformations \ 
\begin{equation}
dq=\sqrt{m\left( x\right) }dx\text{ , }\frac{\partial x}{\partial q}=\frac{1%
}{\sqrt{m\left( x\right) }},
\end{equation}%
and \ $\Psi \left( q,\tau \right) =m\left( x\right) ^{-1/4}\Phi \left(
x,t\right) $ in (18), we immediately get%
\begin{equation*}
\left\langle P_{_{q}}\right\rangle =\int\limits_{-\infty }^{\infty }dx\sqrt{%
m\left( x\right) }\left[ \frac{\Phi ^{\ast }\left( x,t\right) }{m\left(
x\right) ^{1/4}}\right] \left( \frac{-i}{\sqrt{m\left( x\right) }}\frac{%
\partial }{\partial x}\right) \left[ \frac{\Phi \left( x,t\right) }{m\left(
x\right) ^{1/4}}\right] =\int\limits_{-\infty }^{\infty }dx\,\frac{\Phi
^{\ast }\left( x,t\right) }{\sqrt{m\left( x\right) }}\left( -i\left[ \frac{%
\partial }{\partial x}-\frac{1}{4}\left( \frac{\partial _{x}m\left( x\right) 
}{m\left( x\right) }\right) \right] \right) \Phi \left( x,t\right) .
\end{equation*}%
Which clearly suggests that%
\begin{equation}
\widehat{P}\left( q\left( x\right) \right) =\frac{-i}{\sqrt{m\left( x\right) 
}}\left[ \frac{\partial }{\partial x}-\frac{1}{4}\left( \frac{\partial
_{x}m\left( x\right) }{m\left( x\right) }\right) \right]
\end{equation}%
Hereby, one should notice that the construction of this PDM momentum-like
operator (descending from the generalized coordinates settings into
rectangular coordinates settings) has nothings to do with ambiguity
parameters $\alpha ,\beta ,$ and $\gamma $. It is only a manifestation of
mapping the wave functions from one coordinate system to another through
(19). This is not yet the PDM-momentum operator and shall be called PDM
pseudo-mementum operator (with the identity $\widehat{\pi }_{x}\left(
q\left( x\right) \right) $) \cite{49}. The generalization of which is
straightforward and takes the form 
\begin{equation}
\widehat{\pi }_{_{j}}\left( \overrightarrow{q}\left( \vec{x}\right) \right) =%
\frac{-i}{\sqrt{m\left( \vec{x}\right) }}\left[ \frac{\partial }{\partial
x_{_{j}}}-\frac{1}{4}\left( \frac{\partial _{x_{j}}m\left( \vec{x}\right) }{%
m\left( \vec{x}\right) }\right) \right] \Longleftrightarrow \widehat{\pi }%
\left( \overrightarrow{q}\left( \vec{x}\right) \right) =\frac{-i}{\sqrt{%
m\left( \vec{x}\right) }}\left[ \overrightarrow{\nabla }-\frac{1}{4}\left( 
\frac{\overrightarrow{\nabla }m\left( \vec{x}\right) }{m\left( \vec{x}%
\right) }\right) \right] ,
\end{equation}%
where $\widehat{\pi }_{_{j}}\left( \overrightarrow{q}\left( \vec{x}\right)
\right) \rightarrow \widehat{p}_{_{j}}=-i\partial /\partial x_{_{j}}$ for
constant mass settings (i.e., for the dimensionless scalar multiplier $%
m\left( \vec{x}\right) =1$). In fact, equation (21) gives the differential
form of the Hamilton's canonical PDM pseudo-momentum operator $\widehat{\pi }%
\left( \overrightarrow{q}\left( \vec{x}\right) \right) $. Under such
settings, our PDM Schr\"{o}dinger equation (15) inherits the simplistic form%
\begin{equation}
\left\{ \widehat{\pi }_{_{j}}^{2}\left( \overrightarrow{q}\left( \vec{x}%
\right) \right) +V\left( \overrightarrow{q}\left( \vec{x}\right) \right)
\right\} \phi \left( \vec{x}\right) =E\phi \left( \vec{x}\right) .
\end{equation}

Furthermore, one should be aware that for $2m_{\circ }\neq 1$ the first term
of equation (22) would result in $\widehat{\pi }_{_{j}}^{2}\left( 
\overrightarrow{q}\left( \vec{x}\right) \right) /2m_{\circ }$ as the quantum
PDM-kinetic energy operator (i.e., $\widehat{T}=\widehat{\pi }%
_{_{j}}^{2}\left( \overrightarrow{q}\left( \vec{x}\right) \right) /2m_{\circ
}$). Only under such transformation procedure's settings the quantum
Hamiltonian implies the classical one, the other way around holds true as
well. That is, 
\begin{equation}
\widehat{H}_{quantum}=\frac{\widehat{\pi }_{_{j}}^{2}\left( \overrightarrow{q%
}\left( \vec{x}\right) \right) }{2m_{\circ }}+V\left( \overrightarrow{q}%
\left( \vec{x}\right) \right) \Longleftrightarrow H_{classical.}=\frac{1}{2}%
m_{\circ }m\left( \vec{x}\right) \dot{x}_{_{j}}^{2}+V\left( \overrightarrow{q%
}\left( \vec{x}\right) \right) =\frac{\pi _{_{j}}^{2}\left( \overrightarrow{q%
}\left( \vec{x}\right) \right) }{2m_{\circ }}+V\left( \overrightarrow{q}%
\left( \vec{x}\right) \right)
\end{equation}%
where $\pi _{_{j}}\left( \overrightarrow{q}\left( \vec{x}\right) \right) $
is the $j$th-component of the classical PDM pseudo-momentum obtained through 
\begin{equation}
\dot{q}_{_{j}\,}\left( \vec{x}\right) =\sqrt{m\left( \vec{x}\right) }\dot{x}%
_{_{j}}\Longrightarrow \frac{d\vec{q}}{dt}=\dot{q}_{_{j}\,}\widehat{q}%
_{_{j}}=\sqrt{m\left( \vec{x}\right) }\dot{x}_{_{j}\,}\widehat{x}%
_{j}\Longrightarrow \pi _{_{j}}\left( \overrightarrow{q}\left( \vec{x}%
\right) \right) =m_{\circ }\left[ \sqrt{m\left( \vec{x}\right) }\dot{x}%
_{_{j}}\right] ,
\end{equation}%
and $\widehat{\pi }_{_{j}}\left( \overrightarrow{q}\left( \vec{x}\right)
\right) $ is the corresponding $j$th-component of the quantum PDM
pseudo-momentum operator. At this very point, however,\ one recollects the
classical PDM-Lagrangian $L=m_{\circ }m\left( \vec{x}\right) \dot{x}%
_{j}^{2}/2-V\left( \vec{x}\right) $ to imply the classical PDM Hamiltonian $%
H=m_{\circ }m\left( \vec{x}\right) \dot{x}_{j}^{2}/2+V\left( \vec{x}\right)
=P_{_{j}}^{2}/\left[ 2m_{\circ }m\left( \vec{x}\right) \right] +V\left( \vec{%
x}\right) $ where $P_{_{j}}\left( \vec{x}\right) =\partial L/\partial \dot{x}%
_{j}=m_{\circ }m\left( \vec{x}\right) \dot{x}_{j}$ is the canonical
PDM-momentum. This would, in effect, imply that $P_{_{j}}\left( \vec{x}%
\right) =$ $\sqrt{m\left( \vec{x}\right) }\,\pi _{_{j}}\left( 
\overrightarrow{q}\left( \vec{x}\right) \right) $ and consequently the
PDM-momentum operator reads 
\begin{equation}
\widehat{P}_{_{j}}\left( \vec{x}\right) =\sqrt{m\left( \vec{x}\right) }%
\widehat{\pi }_{_{j}}\left( \overrightarrow{q}\left( \vec{x}\right) \right)
=-i\left[ \frac{\partial }{\partial x_{_{j}}}-\frac{1}{4}\left( \frac{%
\partial _{x_{j}}m\left( \vec{x}\right) }{m\left( \vec{x}\right) }\right) %
\right] \Longleftrightarrow \widehat{\pi }_{_{j}}\left( \overrightarrow{q}%
\left( \vec{x}\right) \right) =\frac{\widehat{P}_{_{j}}\left( \vec{x}\right) 
}{\sqrt{m\left( \vec{x}\right) }}
\end{equation}%
where $\widehat{\pi }_{_{j}}\left( \overrightarrow{q}\left( \vec{x}\right)
\right) $ is given in (21). This would necessarily mean that $\widehat{\pi }%
_{_{j}}^{2}\left( \overrightarrow{q}\left( \vec{x}\right) \right) $ of (22)
should be expressed as $\left( \widehat{P}_{_{j}}\left( \vec{x}\right) /%
\sqrt{m\left( \vec{x}\right) }\right) ^{2}$ \ and not as $\widehat{P}%
_{_{j}}^{2}\left( \vec{x}\right) /m\left( \vec{x}\right) $. In classical
mechanics both forms work but not in quantum mechanics. We are now in a
position to dwell on electromagnetic interaction and minimal coupling for
PDM settings.

\section{Classical Electromagnetic interaction and the PDM-quantum
mechanical correspondence}

In this section, we adopt our procedure above and extend it to include
electromagnetic interactions. We begin with the motion of a classical
particle of charge $e$ and a constant rest mass $m_{\circ }$ moving in an
electromagnetic interaction represented by the 4-vector potential $A_{\mu }=$
$\left( \vec{A},i\varphi \right) $ with the vector potential $\vec{A}\left( 
\vec{q}\right) $ and a scalar potential $\varphi (\vec{q})$. The Lagrangian
for such a system is given by%
\begin{equation}
L\left( q_{_{j}},\dot{q}_{_{j}};t\right) =\frac{1}{2}m_{\circ }\dot{q}%
_{_{j}}^{2}+e\,\dot{q}_{_{j}\,}A_{j}\left( \vec{q}\right) -\left[ e\,\varphi
(\vec{q})+V(\vec{q})\right] \text{ };\text{ \ }\dot{q}_{_{j}}=\frac{dq_{_{j}}%
}{dt}.
\end{equation}%
Where $V(\vec{q})$ is any other potential energy than the electric and
magnetic ones. Of course, this is a more general problem than the one
discussed in section II and one may switch-off the electromagnetic
interaction potentials (i.e., $A_{j}\left( \vec{q}\right) =0$ and $\varphi (%
\vec{q})=0$) and, therefore, recover the same results discussed above.

Under such Lagrangian (26) settings, the classical Hamiltonian reads%
\begin{equation}
H\left( q_{_{j}},P_{j};t\right) =\dot{q}_{_{j}\,}P_{j}-L\left( q_{_{j}},\dot{%
q}_{_{j}};t\right) =\frac{1}{2}m_{\circ }\dot{q}_{_{j}}^{2}+W(\vec{q});\text{
\ }W(\vec{q})=e\,\varphi (\vec{q})+V(\vec{q}).
\end{equation}%
Here, the $j$th component of the canonical momentum (associated with the
generalized coordinate $q_{_{j}}$) is given by 
\begin{equation}
P_{j}=\frac{\partial }{\partial \dot{q}_{_{j}\,}}L\left( q_{_{j}},\dot{q}%
_{_{j}};t\right) \Longrightarrow P_{j}=m_{\circ }\dot{q}_{_{j}\,}+eA_{j}%
\left( \vec{q}\right) ,
\end{equation}%
and the classical Hamiltonian (27) takes the form 
\begin{equation}
H\left( q_{_{j}},P_{j};t\right) =\frac{1}{2m_{\circ }}\left(
P_{j}-eA_{j}\left( \vec{q}\right) \right) ^{2}+W(\vec{q}).
\end{equation}%
Hence, the corresponding quantum mechanical Hamiltonian, with $2m_{\circ }=1$
unit and $\widehat{P}_{j}=-i\partial _{q_{_{j}}}$, consequently yields%
\begin{equation}
\hat{H}\left( q_{_{j}},P_{j};t\right) =-\partial _{q_{j}}^{2}+ie\,\left[
\partial _{q_{j}}A_{j}\left( \vec{q}\right) \right] +2ie\,A_{j}\left( \vec{q}%
\right) \,\partial _{q_{j}}+e^{2}A_{j}\left( \vec{q}\right) ^{2}+W(\vec{q}).
\end{equation}%
Now, we follow our methodical proposal in section II above and obtain the
corresponding time-independent PDM-Schr\"{o}dinger equation%
\begin{gather}
\left\{ -\frac{1}{m\left( \vec{x}\right) }\,\partial _{x_{j}}^{2}+\left[ 
\frac{\partial _{x_{j}}m\left( \vec{x}\right) }{m\left( \vec{x}\right) ^{2}}%
\right] \,\partial _{x_{j}}-\frac{7}{16}\left( \frac{\left[ \partial
_{x_{j}}m\left( \vec{x}\right) \right] ^{2}}{m\left( \vec{x}\right) ^{3}}%
\right) +\frac{1}{4}\left[ \frac{\partial _{x_{j}}^{2}m\left( \vec{x}\right) 
}{m\left( \vec{x}\right) ^{2}}\right] +ie\,\frac{\partial
_{x_{j}}A_{j}\left( \vec{q}\left( \vec{x}\right) \right) }{\sqrt{m\left( 
\vec{x}\right) }}\right.  \notag \\
\left. +2ie\,\frac{A_{j}\left( \vec{q}\left( \vec{x}\right) \right) }{\sqrt{%
m\left( \vec{x}\right) }}\left[ \partial _{x_{j}}-\frac{1}{4}\left( \frac{%
\partial _{x_{j}}m\left( \vec{x}\right) }{m\left( \vec{x}\right) }\right) %
\right] +e^{2}A_{j}\left( \vec{q}\left( \vec{x}\right) \right) ^{2}+W\left( 
\vec{q}\left( \vec{x}\right) \right) \right\} \phi \left( \vec{x}\right)
=E\phi \left( \vec{x}\right) ,
\end{gather}%
that reduces into%
\begin{equation}
\left\{ \left[ \frac{-i}{\sqrt{m\left( \vec{x}\right) }}\left[ \frac{%
\partial }{\partial x_{_{j}}}-\frac{1}{4}\left( \frac{\partial
_{x_{j}}m\left( \vec{x}\right) }{m\left( \vec{x}\right) }\right) \right]
-e\,A_{j}\left( \vec{q}\left( \vec{x}\right) \right) \right] ^{2}+W(\vec{q}%
\left( \vec{x}\right) )\right\} \phi \left( \vec{x}\right) =E\phi \left( 
\vec{x}\right) .
\end{equation}%
or in a more simplistic format%
\begin{equation}
\left\{ \left[ \frac{\widehat{P}_{_{j}}\left( \vec{x}\right) }{\sqrt{m\left( 
\vec{x}\right) }}-e\,A_{j}\left( \vec{q}\left( \vec{x}\right) \right) \right]
^{2}+W(\vec{q}\left( \vec{x}\right) )\right\} \phi \left( \vec{x}\right)
=E\phi \left( \vec{x}\right) .
\end{equation}%
where the scalar potential $W(\vec{q}\left( \vec{x}\right) )=W\left( \vec{x}%
\right) $ and the vector potential $A_{j}\left( \vec{q}\left( \vec{x}\right)
\right) $ is yet to be correlated with $A_{j}\left( \vec{x}\right) $ in the
sequel.

Classical mechanically, the PDM-Lagrangian and PDM-Hamiltonian with
electromagnetic interaction are of the forms%
\begin{equation}
L=\frac{1}{2}m_{\circ }m\left( \vec{x}\right) \dot{x}_{j}^{2}+e\,\dot{x}%
_{j}A_{j}\left( \vec{x}\right) -W\left( \vec{x}\right) \Longleftrightarrow H=%
\frac{1}{2}m_{\circ }m\left( \vec{x}\right) \dot{x}_{j}^{2}+W\left( \vec{x}%
\right) ;\text{ \ }W\left( \vec{x}\right) =e\,\varphi (\vec{x})+V(\vec{x}),
\end{equation}%
where the PDM-canonical momentum reads%
\begin{equation}
P_{_{j}}\left( \vec{x}\right) =\frac{\partial L}{\partial \dot{x}_{j}}%
=m_{\circ }m\left( \vec{x}\right) \dot{x}_{j}+eA_{j}\left( \vec{x}\right)
\Longleftrightarrow m_{\circ }\sqrt{m\left( \vec{x}\right) }\dot{x}_{j}=%
\frac{P_{_{j}}\left( \vec{x}\right) -eA_{j}\left( \vec{x}\right) }{\sqrt{%
m\left( \vec{x}\right) }}.
\end{equation}%
Therefore, in terms of the canonical momentum the PDM-Hamiltonian (34) takes
the form%
\begin{equation}
H=\frac{1}{2m_{\circ }}\left( \frac{P_{_{j}}\left( \vec{x}\right)
-eA_{j}\left( \vec{x}\right) }{\sqrt{m\left( \vec{x}\right) }}\right)
^{2}+W\left( \vec{x}\right) ,
\end{equation}%
and the quantum mechanical PDM-Hamiltonian hence reads (in $\hbar =2m_{\circ
}=1$ units)%
\begin{equation}
\widehat{H}=\left( \frac{\widehat{P}_{_{j}}\left( \vec{x}\right)
-eA_{j}\left( \vec{x}\right) }{\sqrt{m\left( \vec{x}\right) }}\right)
^{2}+W\left( \vec{x}\right) .
\end{equation}%
Which immediately, when compared with the PDM-Hamiltonian of (33), suggests
the correlation between the vector potentials $A_{j}\left( \vec{q}\left( 
\vec{x}\right) \right) $ and $A_{j}\left( \vec{x}\right) $ as 
\begin{equation}
A_{j}\left( \vec{q}\left( \vec{x}\right) \right) =\frac{A_{j}\left( \vec{x}%
\right) }{\sqrt{m\left( \vec{x}\right) }}.
\end{equation}%
Consequently equation (33) should look like%
\begin{equation}
\left\{ \left( \frac{\widehat{P}_{_{j}}\left( \vec{x}\right) -eA_{j}\left( 
\vec{x}\right) }{\sqrt{m\left( \vec{x}\right) }}\right) ^{2}+W\left( \vec{x}%
\right) \right\} \phi \left( \vec{x}\right) =E\phi \left( \vec{x}\right) .
\end{equation}

It now obvious, therefore, that the simplest way of coupling the
electromagnetic interaction is to take the Hamilton's canonical
pseudo-momentum $\pi _{_{j}}\left( \vec{q}\left( \vec{x}\right) \right) $ as
the sum of the kinetic momentum $m_{\circ }\dot{q}_{_{j}\,}=m_{\circ }\left( 
\sqrt{m\left( \vec{x}\right) }\dot{x}_{_{j}}\right) $ and $eA_{j}\left( \vec{%
q}\left( \vec{x}\right) \right) $ ( i.e., $\pi _{_{j}}\left( \vec{q}\left( 
\vec{x}\right) \right) =m_{\circ }\left( \sqrt{m\left( \vec{x}\right) }\dot{x%
}_{_{j}}\right) +eA_{j}\left( \vec{q}\left( \vec{x}\right) \right) $).
Hence, for the classical Hamiltonian in (23) one may simply use the minimal
coupling 
\begin{equation*}
\pi _{_{j}}\left( \vec{q}\left( \vec{x}\right) \right) =\left( \frac{%
P_{_{j}}\left( \vec{x}\right) }{\sqrt{m\left( \vec{x}\right) }}\right)
\longrightarrow \pi _{_{j}}\left( \vec{q}\left( \vec{x}\right) \right)
-eA_{j}\left( \vec{q}\left( \vec{x}\right) \right) \text{ \ and \ }%
E=H_{classical}\longrightarrow E-e\,\varphi (\vec{q}\left( \vec{x}\right) ),
\end{equation*}%
or in terms of the canonical PDM-momentum, it precisely reads%
\begin{equation}
\left( \frac{P_{_{j}}\left( \vec{x}\right) }{\sqrt{m\left( \vec{x}\right) }}%
\right) \longrightarrow \left( \frac{P_{_{j}}\left( \vec{x}\right)
-eA_{j}\left( \vec{x}\right) }{\sqrt{m\left( \vec{x}\right) }}\right) \text{
\ and \ }E=H_{classical}\longrightarrow E-e\,\varphi (\vec{q}\left( \vec{x}%
\right) ),
\end{equation}%
to incorporate electromagnetic interactions. Consequently, in quantum
mechanics, it is obvious that the electromagnetic interactions for PDM are
integrated into the PDM-Schr\"{o}dinger equation (22) through the the
minimal coupling 
\begin{equation}
\widehat{\pi }_{_{j}}\left( \vec{q}\left( \vec{x}\right) \right) =\frac{%
\widehat{P}_{_{j}}\left( \vec{x}\right) }{\sqrt{m\left( \vec{x}\right) }}%
\longrightarrow \left( \frac{\widehat{P}_{_{j}}\left( \vec{x}\right)
-eA_{j}\left( \vec{x}\right) }{\sqrt{m\left( \vec{x}\right) }}\right) \text{
\ \ and \ \ }E\longrightarrow E-e\,\varphi (\vec{q}\left( \vec{x}\right) ).
\end{equation}%
Which, in fact, looks very much like the usual constant mass settings but
now with the Classical PDM momentum $P_{_{j}}\left( \vec{x}\right) =\sqrt{%
m\left( \vec{x}\right) }\pi _{_{j}}\left( \vec{q}\left( \vec{x}\right)
\right) $ of (24) and the Quantum PDM momentum operator $\widehat{P}%
_{_{j}}\left( \vec{x}\right) $ of (25) rather than the textbook momentum $%
p_{_{j}}=m_{\circ }\dot{x}_{_{j}}$ and momentum operator $\widehat{p}%
_{_{j}}=-i\partial /\partial x$, respectively. This result renders the
procedure followed by Dutra and Oliveira \cite{10} inappropriate, for they
have started with their equation (37) assuming that $\widehat{p}%
_{_{j}}=-i\partial /\partial x_{_{j}}$ and $\widehat{p}_{_{j}}%
\longrightarrow \widehat{p}_{_{j}}-e\,A_{_{j}}\left( \vec{x}\right) $ (this
recipe is correct only for constant mass settings). However, this readily
lies far beyond our methodical proposal and shall be discussed elsewhere.

Nevertheless, one should notice that a proper reverse engineering of (39),
with $\phi \left( \vec{x}\right) =m\left( \vec{x}\right) ^{1/4}\psi \left( 
\vec{q}\right) $, would immediately yield%
\begin{equation}
\left\{ \left[ \widehat{P}_{j}-eA_{j}\left( \vec{q}\right) \right] ^{2}+W(%
\vec{q})\right\} \psi \left( \vec{q}\right) =E\psi \left( \vec{q}\right) ;\,%
\text{\ \ }\widehat{P}_{j}=-i\partial _{q_{j}}
\end{equation}%
Which clearly introduces a paramagnetic contribution as $2ie\,A_{j}\left( 
\vec{q}\right) \,\partial _{q_{j}}$ and a diamagnetic one as $%
e^{2}A_{j}\left( \vec{q}\right) ^{2}$ along with an electric field
contribution as $e\,\varphi (\vec{q})$ in the generalized coordinates.
Obviously, equation (42) represents a textbook example which is known to be
exactly or conditionally exactly solvable model for some $W(\vec{q})$ forms.
The solutions of which can be mapped into the PDM Schr\"{o}dinger equation
(39).

\section{Eligibility of the vector potentials and PDM-settings}

In this section, we shall consider the two vector potentials that satisfy
the Coulomb gauge $\partial _{q_{j}}A_{j}\left( \vec{q}\right) =0$ and are
often used in the literature as illustrative examples. They are,%
\begin{equation}
\vec{A}\left( \vec{q}\right) =B_{\circ }(-q_{_{2}},0,0)=-B_{\circ }q_{_{2}}\,%
\widehat{q}_{_{1}}
\end{equation}%
and 
\begin{equation}
\vec{A}\left( \vec{q}\right) =\frac{B_{\circ }}{2}(-q_{_{2}},q_{_{1}},0)=%
\frac{B_{\circ }}{2}\left( -q_{_{2}}\,\widehat{q}_{_{1}}+q_{_{1}}\,\widehat{q%
}_{_{2}}\right)
\end{equation}%
where $\widehat{q}_{_{i}}$ is the unit vector for the generalized coordinate 
$q_{_{i}}$. Consequently, they result a constant magnetic field 
\begin{equation}
\vec{B}\left( \vec{q}\right) =\vec{\nabla}_{q}\times \vec{A}\left( \vec{q}%
\right) =\left\{ 
\begin{tabular}{l}
$B_{\circ }\,\widehat{q}_{_{3}}$ for $\vec{A}\left( \vec{q}\right) =B_{\circ
}(-q_{_{2}},0,0)\mathstrut \medskip $ \\ 
$\frac{B_{\circ }}{2}\,\widehat{q}_{_{3}}$ for $\vec{A}\left( \vec{q}\right)
=\frac{B_{\circ }}{2}(-q_{_{2}},q_{_{1}},0)\medskip \mathstrut $%
\end{tabular}%
\right. ;\text{ }\vec{\nabla}_{q}=\widehat{q}_{_{1}}\partial _{q_{1}}+%
\widehat{q}_{_{2}}\partial _{q_{2}}+\widehat{q}_{_{3}}\partial _{q_{3}}
\end{equation}%
Hereby, we shall subject the two vector potentials $\vec{A}\left( \vec{q}%
\right) $ in (43) and (44) to some eligibility test in order to be able to
deal with Schr\"{o}dinger equation (42) for different interaction potentials
(be it the vector potentials $A_{j}\left( \vec{q}\right) $ and/or scalar
potentials $W(\vec{q})=e\,\varphi (\vec{q})+V(\vec{q})$) and hence to\
reflect on the corresponding PDM settings in (39).

A priori, however, let us recollect the correlation of (38) and recast it as%
\begin{equation}
A_{j}\left( \vec{q}\left( \vec{x}\right) \right) =\frac{A_{j}\left( \vec{x}%
\right) }{\sqrt{m\left( \vec{x}\right) }}\Longrightarrow A_{j}\left( \vec{q}%
\left( \vec{x}\right) \right) =\frac{S\left( \vec{x}\right) }{\sqrt{m\left( 
\vec{x}\right) }}\tilde{A}_{_{j}}\left( \vec{x}\right) \text{ };\,\text{\ }%
\overrightarrow{\tilde{A}}\left( \vec{x}\right) =\left\{ 
\begin{tabular}{l}
$B_{\circ }(-x_{_{2}},0,0)\mathstrut \medskip $ \\ 
$\frac{B_{\circ }}{2}(-x_{_{2}},x_{_{1}},0)$%
\end{tabular}%
\right. ,
\end{equation}%
where the introduction of the scalar multiplier $S\left( \vec{x}\right) $ in
the assumption $A_{j}\left( \vec{x}\right) =S\left( \vec{x}\right) \tilde{A}%
_{_{j}}\left( \vec{x}\right) $ absorbs any other position-dependent terms
that may emerge from the construction of the vector potential $%
\overrightarrow{A}\left( \vec{x}\right) $ (such as that of a long solenoid
for example). Yet, the Coulomb gauge $\partial _{q_{j}}A_{j}\left( \vec{q}%
\right) =0$ should be satisfied and remain invariant under our point
transformation. That is, with $m\left( \vec{x}\right) =m\left( r\right) $, $%
S\left( \vec{x}\right) =S\left( r\right) $ and $S^{\prime }\left( r\right)
=dS\left( r\right) /dr$ $;\,r=\sqrt{x_{_{1}}^{2}+x_{_{2}}^{2}+x_{_{3}}^{2}}$%
, the condition%
\begin{equation}
\partial _{q_{j}}A_{j}\left( \vec{q}\left( \vec{x}\right) \right) =\frac{1}{%
\sqrt{m\left( r\right) }}\partial _{x_{j}}\left( \frac{S\left( r\right) }{%
\sqrt{m\left( r\right) }}\tilde{A}_{j}\left( \vec{x}\right) \right) =\frac{%
S\left( r\right) }{m\left( r\right) }\left[ \,\partial _{x_{j}}\tilde{A}%
_{j}\left( \vec{x}\right) +\frac{\,x_{_{j}}\tilde{A}_{j}\left( \vec{x}%
\right) }{r}\left( \frac{S^{\prime }\left( r\right) }{S\left( r\right) }-%
\frac{m^{\prime }\left( r\right) }{2m\left( r\right) }\right) \right] =0.
\end{equation}%
has to be satisfied. It is obvious that, whilst the first term\ $\,\partial
_{x_{j}}\tilde{A}_{j}\left( \vec{x}\right) =0$ for both forms of $%
\overrightarrow{\tilde{A}}\left( \vec{x}\right) $ in (46), the second term\ $%
x_{_{j}}\tilde{A}_{j}\left( \vec{x}\right) =0$ if and only if $%
\overrightarrow{\tilde{A}}\left( \vec{x}\right) =\frac{B_{\circ }}{2}%
(-x_{_{2}},x_{_{1}},0)$. We, therefore, consider the only eligible vector
potential setting 
\begin{equation}
\vec{A}\left( \vec{q}\right) =\frac{B_{\circ }}{2}(-q_{_{2}},q_{_{1}},0)=%
\frac{S\left( r\right) }{\sqrt{m\left( r\right) }}\overrightarrow{\tilde{A}}%
\left( \vec{x}\right) \Longleftrightarrow \frac{B_{\circ }}{2}%
(-q_{_{2}},q_{_{1}},0)=\frac{S\left( r\right) }{\sqrt{m\left( r\right) }}%
\frac{B_{\circ }}{2}(-x_{_{2}},x_{_{1}},0).
\end{equation}%
This would immediately imply that%
\begin{equation}
q_{_{j}}\left( \vec{x}\right) =\frac{S\left( r\right) }{\sqrt{m\left(
r\right) }}\,x_{_{j}}\Longrightarrow \frac{\partial q_{_{j}}}{\partial
x_{_{j}}}=\frac{S\left( r\right) }{\sqrt{m\left( r\right) }}\left[ 1+\frac{%
x_{_{j}}^{2}}{r}\left( \frac{S^{\prime }\left( r\right) }{S\left( r\right) }-%
\frac{m^{\prime }\left( r\right) }{2m\left( r\right) }\right) \right] ,\text{
(no summation)}.\text{ }
\end{equation}%
Which when summed over the repeated index yields%
\begin{equation}
\frac{\partial q_{_{j}}}{\partial x_{_{j}}}=\frac{S\left( r\right) }{\sqrt{%
m\left( r\right) }}\left[ N+r\left( \frac{S^{\prime }\left( r\right) }{%
S\left( r\right) }-\frac{m^{\prime }\left( r\right) }{2m\left( r\right) }%
\right) \right] ,
\end{equation}%
where $N\geq 2$ \ denotes the number of degrees of freedom involved in the
problem at hand and in our case $N=3$. If we now use equation (9) (with $%
g\left( \vec{x}\right) =m\left( \vec{x}\right) =m(r)$) summed up over the
repeated index, we get%
\begin{equation}
\frac{\partial q_{_{j}}}{\partial x_{_{j}}}=N\sqrt{m\left( \vec{x}\right) }=N%
\sqrt{m\left( r\right) }
\end{equation}%
Hence, (50) and (51) suggest the relation%
\begin{equation}
m\left( r\right) =S\left( r\right) \left[ 1+\frac{r}{N}\left( \frac{%
S^{\prime }\left( r\right) }{S\left( r\right) }-\frac{m^{\prime }\left(
r\right) }{2m\left( r\right) }\right) \right] \Longleftrightarrow S\left(
r\right) =N\sqrt{m\left( r\right) }r^{-N}\int r^{N-1}\sqrt{m\left( r\right) }%
dr.
\end{equation}%
Although $N=3$ for the current methodical proposal, we choose to cast the
above equation in terms of $N$ to identify the number of degrees of freedom
involved in the problem at hand. Moreover, for a given $m\left( r\right) $
one may find $S\left( r\right) $ using (52), the other way around works as
well. Therefore, $m\left( r\right) $ and $S\left( r\right) $ may very well
be considered as generating functions of each other. That is, one may start
with $S\left( r\right) $ to find $m\left( r\right) $, for example, 
\begin{eqnarray}
S\left( r\right) &=&1\Longleftrightarrow m\left( r\right) =\frac{1}{%
1+\lambda r^{-2N}}, \\
S\left( r\right) &=&m\left( r\right) \Longleftrightarrow m\left( r\right)
=const.,\medskip \\
S\left( r\right) &=&m\left( r\right) ^{b}\Longleftrightarrow m\left(
r\right) =\left[ 1+\lambda r^{-2N\left( b-1\right) /\left( 2b-1\right) }%
\right] ^{1/\left( b-1\right) };\,b\neq 1\medskip , \\
S\left( r\right) &=&\lambda r^{\nu }\Longleftrightarrow m\left( r\right) =%
\frac{\left( 2N+\nu \right) \lambda }{\left( 2N+\nu \right) \lambda
r^{-2\left( N+\nu \right) }+2Nr^{-\nu }},
\end{eqnarray}%
or one starts with $m\left( r\right) $ to find $S\left( r\right) $ 
\begin{eqnarray}
m\left( r\right) &=&\lambda r^{2}\iff S\left( r\right) =\frac{N\lambda r^{2}%
}{N+1}, \\
m\left( r\right) &=&\lambda r^{2b}\iff S\left( r\right) =\frac{N\lambda
r^{2b}}{N+b}\medskip , \\
m\left( r\right) &=&\frac{1}{1+\alpha r^{N}}\Longleftrightarrow S\left(
r\right) =\frac{2}{\alpha }r^{-N},
\end{eqnarray}%
and so on so forth. In what follows, we clarify our methodical proposal.

\section{Illustrative Examples}

\subsection{Charged PDM-particle in a vector potential $\vec{A}\left( \vec{q}%
\right) =\frac{B_{\circ }}{2}(-q_{_{2}},q_{_{1}},0)$ and $W(\vec{q})=0$}

A charged particle moving under the influence of the vector potential $\vec{A%
}\left( \vec{q}\right) =B_{\circ }(-q_{_{2}},q_{_{1}},0)/2\Longrightarrow 
\vec{B}\left( \vec{q}\right) =B_{\circ }\widehat{q}_{_{3}}/2=B_{\circ }%
\widehat{x}_{_{3}}/2$ would be described by the Schr\"{o}dinger equation
(42) as 
\begin{equation}
\left\{ \left[ \widehat{P}_{_{1}}+\frac{eB_{\circ }}{2}q_{_{2}}\right] ^{2}+%
\left[ \widehat{P}_{_{2}}-\frac{eB_{\circ }}{2}q_{_{1}}\right] ^{2}+\widehat{%
P}_{_{3}}^{2}\right\} \psi \left( \vec{q}\right) =E\psi \left( \vec{q}%
\right) .
\end{equation}%
This would, in effect, suggest that the Hamiltonian 
\begin{equation*}
\widehat{H}\left( q,P\right) =\left[ \widehat{P}_{_{1}}+eB_{\circ }q_{_{2}}%
\right] ^{2}+\left[ \widehat{P}_{_{2}}-\frac{eB_{\circ }}{2}q_{_{1}}\right]
^{2}+\widehat{P}_{3}^{2}
\end{equation*}%
does not explicitly depend on $q_{_{3}}$, and the commutation relations%
\begin{equation}
\left[ q_{_{i}},\widehat{P}_{_{j}}\right] =i\delta _{ij}\text{ \ , \ }\left[ 
\widehat{P}_{i},\widehat{P}_{_{j}}\right] =0\text{ \ , \ }\left[ \widehat{P}%
_{_{3}},\widehat{H}\left( q,P\right) \right] =0
\end{equation}%
are satisfied. Hence, $\widehat{P}_{_{3}}$ is no longer an operator but
rather\ a constants of motion (i.e., it can be replaced by a number,
therefore). Consequently, the solution of (60) can be expressed as%
\begin{equation}
\psi \left( \vec{q}\right) =\exp \left[ i\left(
k_{_{1}}q_{_{1}}+k_{_{3}}q_{_{3}}+\frac{eB_{\circ }}{2}q_{_{1}}q_{_{2}}%
\right) \right] Y\left( q_{_{2}}\right)
\end{equation}%
to result in a shifted harmonic oscillator like Schr\"{o}dinger equation%
\begin{equation}
\left\{ -\frac{d^{2}}{dq_{_{2}}^{2}}+e^{2}B_{\circ }^{2}\left[ q_{_{2}}+%
\frac{k_{_{1}}}{eB_{\circ }}\right] ^{2}+k_{_{3}}^{2}\right\} Y_{n}\left(
q_{_{2}}\right) =E_{n}Y_{n}\left( q_{_{2}}\right) .
\end{equation}%
Which admits exact energy eigenvalues and eigenfunctions, respectively, as%
\begin{eqnarray}
E_{n} &=&k_{3}^{2}+\left( 2n+1\right) \left\vert e\right\vert B_{\circ },%
\text{ \mathstrut \medskip \medskip \medskip } \\
Y_{n}\left( \zeta \right) &\sim &\exp \left[ -\frac{\left\vert e\right\vert
B_{\circ }}{2}\zeta ^{2}\right] H_{n}\left( \sqrt{\left\vert e\right\vert
B_{\circ }}\zeta \right) ;\text{ \ }\zeta =q_{_{2}}+\frac{k_{1}}{eB_{\circ }}%
,\text{ }n=0,1,2,\cdots ,
\end{eqnarray}%
where $H_{n}\left( x\right) $ are the Hermite polynomials

\subsection{Charged PDM-particle in a vector potential $\vec{A}\left( \vec{q}%
\right) =\frac{B_{\circ }}{2}(-q_{_{2}},q_{_{1}},0)$ and $W(\vec{q})=-e%
\mathcal{E}_{\circ }q_{_{2}}$}

Here we take the same charged particle as above and subject it not only to a
constant magnetic field but also to a constant electric field $%
\overrightarrow{E}=\mathcal{E}_{\circ }\widehat{q}_{_{2}}$ (i.e., $\ 
\overrightarrow{E}=\mathcal{E}_{\circ }\widehat{x}_{_{2}};\,\ \widehat{q}%
_{_{2}}=$ $\widehat{x}_{_{2}}$). In this case, our Schr\"{o}dinger equation
(42) reads%
\begin{equation}
\left\{ \left[ \widehat{P}_{_{1}}+\frac{eB_{\circ }}{2}q_{_{2}}\right] ^{2}+%
\left[ \widehat{P}_{_{2}}-\frac{eB_{\circ }}{2}q_{_{1}}\right] ^{2}+\widehat{%
P}_{_{3}}^{2}-e\mathcal{E}_{\circ }q_{_{2}}\right\} \psi \left( \vec{q}%
\right) =E\psi \left( \vec{q}\right) .
\end{equation}%
with the substitution of $\psi \left( \vec{q}\right) $ in (62) we obtain,
again, a shifted harmonic oscillator like Schr\"{o}dinger equation 
\begin{equation}
\left\{ -\frac{d^{2}}{dq_{_{2}}^{2}}+e^{2}B_{\circ }^{2}\left[
q_{_{2}}+\left( \frac{k_{_{1}}}{eB_{\circ }}-\frac{\mathcal{E}_{\circ }}{%
2eB_{\circ }^{2}}\right) \right] ^{2}+k_{_{3}}^{2}\right\} Y_{n}\left(
q_{_{2}}\right) =\tilde{E}_{n}Y_{n}\left( q_{_{2}}\right) .
\end{equation}%
Which admits exact solution similar to that of (63) where the exact
eigenenergies are given by%
\begin{equation}
\tilde{E}_{n}=k_{3}^{2}+\left( 2n+1\right) \left\vert e\right\vert B_{\circ
}\Longrightarrow E_{n}=\left( 2n+1\right) \left\vert e\right\vert B_{\circ
}+k_{_{3}}^{2}+\frac{k_{_{1}}\mathcal{E}_{\circ }}{B_{\circ }}-\frac{%
\mathcal{E}_{\circ }^{2}}{4B_{\circ }^{2}},
\end{equation}%
and the exact eigenfunctions are%
\begin{equation}
Y_{n}\left( \eta \right) \sim \exp \left[ -\frac{\left\vert e\right\vert
B_{\circ }}{2}\eta ^{2}\right] H_{n}\left( \sqrt{\left\vert e\right\vert
B_{\circ }}\eta \right) ;\ \eta =q_{_{2}}+\left( \frac{k_{_{1}}}{eB_{\circ }}%
-\frac{\mathcal{E}_{\circ }}{2eB_{\circ }^{2}}\right) ,n=0,1,2,\cdots .
\end{equation}

For both illustrative examples above, one may recollect our coordinates'
settings of (9)-(11) and (49), along with $g\left( \vec{x}\right) =m\left( 
\vec{x}\right) =m\left( r\right) $, to come out with%
\begin{equation*}
\vec{A}\left( \vec{q}\left( \vec{x}\right) \right) =\frac{\vec{A}\left( \vec{%
x}\right) }{\sqrt{m\left( r\right) }}=\frac{S\left( r\right) }{\sqrt{m\left(
r\right) }}\overrightarrow{\tilde{A}}\left( \vec{x}\right) =\frac{S\left(
r\right) }{\sqrt{m\left( r\right) }}\frac{B_{\circ }}{2}%
(-x_{_{2}},x_{_{1}},0)\,\Longrightarrow \text{ \ }q_{_{j}}\left( \vec{x}%
\right) =\frac{S\left( r\right) }{\sqrt{m\left( r\right) }}\,x_{_{j}},
\end{equation*}%
in order to build up the wavefunctions in the rectangular coordinates using $%
\phi \left( \vec{x}\right) =m\left( r\right) ^{1/4}\psi \left( \vec{q}\left( 
\vec{x}\right) \right) $ of (14) and a scalar multiplier $m\left( r\right) $
as those exemplified in (53)-(59), excluding that of (54) for it leads to
constant mass settings that need not be reiterated here again. Hereby, we
notice that all such PDM functions satisfying (52) share the same
eigenvalues and eigenfunctions of either (64) and (65) or (68) and (69),
respectively. Isospectrality is an obvious consequence of the current
methodical proposal, of course.

\section{Concluding Remarks}

In this paper, we have started with the PDM Lagrangian (3) for a classical
particle of mass $m_{\circ }$ moving, in the generalized coordinates $\vec{q}%
=\left( q_{_{1}},q_{_{2}},q_{_{3}}\right) $, under the influence of a scalar
potential field $V\left( \vec{q}\right) $ and built up the corresponding
classical Hamiltonian (6) as well as the quantum Hamiltonian operator (7).
We have shown that the correlation between the associated time-independent
Schr\"{o}dinger equation (8), in the generalized coordinates,\ and the well
known time-independent PDM Schr\"{o}dinger equation (2) is an obvious
consequence of a particular type of point transformation (9)-(12) along with
the transformation of the wave function (14). Using no other constraint than
the von Roos one $\alpha +\beta +\gamma =-1$, it turned out that the
ordering ambiguity in the von Roos Hamiltonian (1), vanishes as a result of
comparison between the second terms of equation (2) and that of (13). The
ordering parameters are strictly determined as $\beta =-1/2$, $\alpha =-1/4$%
, and $\gamma =-1/4$ (documented in (14)-(16) and known in the literature as
MM-ordering of Mustafa and Mazharimousavi \cite{49,50}). Yet, we were able
to use the correlation between the wave functions ( i.e. $\psi \left( \vec{q}%
\right) =m\left( \vec{x}\right) ^{-1/4}\phi \left( \vec{x}\right) $) and
construct the PDM pseudo-momentum operator in one dimension (see (17) -
(20)) and generalize it to three dimensions in (21). To the best of our
knowledge, the construction of the PDM-momentum operator as 
\begin{equation}
\widehat{P}_{_{j}}\left( \vec{x}\right) =\sqrt{m\left( \vec{x}\right) }%
\widehat{\pi }_{_{j}}\left( \vec{q}\left( \vec{x}\right) \right) =-i\left[ 
\frac{\partial }{\partial x_{_{j}}}-\frac{1}{4}\left( \frac{\partial
_{x_{j}}m\left( \vec{x}\right) }{m\left( \vec{x}\right) }\right) \right]
\end{equation}%
has never been reported elsewhere in the literature. Therefore, it is
necessary and vital to cast the PDM-Schr\"{o}dinger equation as%
\begin{equation}
\left\{ \left( \frac{\widehat{P}_{_{j}}\left( \vec{x}\right) }{\sqrt{m\left( 
\vec{x}\right) }}\right) ^{2}+V\left( \vec{x}\right) \right\} \phi \left( 
\vec{x}\right) =E\phi \left( \vec{x}\right) .
\end{equation}%
This would, in effect, fix the ordering ambiguity problem in the von Roos
Hamiltonian (1) that has been known in the literature for few decades.
Therefore, only under our especial type of point transformation settings
that\ the PDM classical and quantum mechanical correspondence\ as well as
the construction of the PDM-momentum (operator) were made feasible.

On the electromagnetic interactions side, moreover, it was very vital to
start again (in section IV) from classical mechanics to find out the
corresponding quantum mechanical settings and dwell on the nature of the
minimal coupling. We have observed that for the classical\ PDM-Hamiltonian
in (23) one may simply use the usual textbook minimal coupling 
\begin{equation}
\pi _{_{j}}\left( \vec{q}\left( \vec{x}\right) \right) \longrightarrow \pi
_{_{j}}\left( \vec{q}\left( \vec{x}\right) \right) -e\,A_{j}\left( \vec{q}%
\left( \vec{x}\right) \right) \text{ }\Longrightarrow P_{_{j}}\left( \vec{x}%
\right) \longrightarrow P_{_{j}}\left( \vec{x}\right) -e\,A_{j}\left( \vec{x}%
\right) .
\end{equation}%
Whereas, for the quantum PDM-Hamiltonian in (22) (equivalently, in the
PDM-Schr\"{o}dinger equation (71)), it is necessary and vital to use the
unusual minimal coupling as 
\begin{equation}
\widehat{\pi }_{_{j}}\left( \vec{q}\left( \vec{x}\right) \right)
\longrightarrow \widehat{\pi }_{_{j}}\left( \vec{q}\left( \vec{x}\right)
\right) -e\,A_{j}\left( \vec{q}\left( \vec{x}\right) \right) \Longrightarrow 
\frac{\widehat{P}_{_{j}}\left( \vec{x}\right) }{\sqrt{m\left( \vec{x}\right) 
}}\longrightarrow \frac{\left( \widehat{P}_{_{j}}\left( \vec{x}\right)
-e\,A_{j}\left( \vec{x}\right) \right) }{\sqrt{m\left( \vec{x}\right) }},
\end{equation}%
where $\widehat{P}_{_{j}}\left( \vec{x}\right) $ is the $j$th component of
the PDM-momentum operator (25). Of course, such unusual minimal coupling
(73) does not hold true for the Klein-Gordon and Dirac relativistic
operators. The PDM concept fits very well without any ambiguity conflict
into Klein-Gordon and Dirac equations (see \cite{52,53,54} for more details
on this issue). Such a PDM-minimal coupling formation (73) has never been
reported elsewhere in the literature, to the best of our knowledge, .

Furthermore, among the two commonly used vector potentials that satisfy the
Coulomb gauge $\partial _{q_{j}}A_{j}\left( \vec{q}\right) =0$ in the
generalized coordinates (i.e., $\vec{A}\left( \vec{q}\left( \vec{x}\right)
\right) =B_{\circ }(-q_{_{2}}\left( \vec{x}\right) ,0,0)$ and $\vec{A}\left( 
\vec{q}\left( \vec{x}\right) \right) =B_{\circ }(-q_{_{2}}\left( \vec{x}%
\right) ,q_{_{1}}\left( \vec{x}\right) ,0)/2$), we found out that the vector
potential%
\begin{equation}
\vec{A}\left( \vec{q}\left( \vec{x}\right) \right) =\frac{B_{\circ }}{2}%
(-q_{_{2}}\left( \vec{x}\right) ,q_{_{1}}\left( \vec{x}\right) ,0)=\frac{%
S\left( r\right) }{\sqrt{m\left( r\right) }}\frac{B_{\circ }}{2}%
(-x_{_{2}},x_{_{1}},0).
\end{equation}%
is the only eligible one that satisfies\ the Coulomb gauge $\partial
_{q_{j}}A_{j}\left( \vec{q}\right) =0$ (within our PDM-point transformation
settings, of course). Two illustrative examples that include magnetic and
electric fields are used, and the mapping between the constant mass settings
and PDM settings is made clear (in section VI).

Finally, although our methodical proposal above is introduced to deal with a
three-dimensional PDM-Schr\"{o}dinger equation, it is also feasibly
applicable to a more commonly used two-dimensional problems (c.f., e.g.
Dutra and Oliveira \cite{10} or Correa et al. \cite{2} and related
references therein). However, the three-dimensional case is a more general
and instructive one.\newpage\


\begin{thebibliography}{99}
\bibitem{1} O. von Roos, Phys. Rev. \textbf{B 27 }(1983) 7547.

\bibitem{2} A. de Souza Dutra, C A S Almeida, Phys Lett. \textbf{A 275}
(2000) 25.

R.A.C. Correa, A. de Souza Dutra, J A de Oliveira, M.G. Garcia, J. Math.
Phys. \textbf{58} (2017) 012104..

\bibitem{3} M. V. Loffe, D. N. Nishnianidze. V.V. Vereshagin, J. Math. Phys. 
\textbf{58} (2017) 072105.

\bibitem{4} O. Mustafa, S. H. Mazharimousavi, Phys. Lett. \textbf{A 358}
(2006) 259.

\bibitem{5} A. D. Alhaidari, Phys. Rev. \textbf{A 66} (2002) 042116.

\bibitem{6} B. Bagchi, A. Banerjee, C. Quesne, V. M. Tkachuk, J. \ Phys. 
\textbf{A}: Math. Gen. \textbf{38} (2005) 2929.

\bibitem{7} O. Mustafa, S. H. Mazharimousavi, Phys. Lett. \textbf{A 357}
(2006) 295.

\bibitem{8} O. Mustafa, J Phys \textbf{A}: Math. Theor. \textbf{44 (}2011%
\textbf{) }355303.

\bibitem{9} B. Bagchi, P. Gorain, C. Quesne and R. Roychoudhury, Mod. Phys.
Lett. \textbf{A 19} (2004) 2765.

\bibitem{10} A . de Souza Dutra, \ J A de Oliveira, J. Phys. \textbf{A}:
Math. Theor. \textbf{42,} (2009) 025304.

\bibitem{11} S. Cruz y Cruz, O Rosas-Ortiz, J Phys \textbf{A}: Math. Theor. 
\textbf{42} (2009) 185205.

\bibitem{12} S. Cruz y Cruz, J. Negro and L.M. Nieto, Phys. Lett. \textbf{A} 
\textbf{369,} (2007) 400.

\bibitem{13} R. Koc, G. Sahinoglu, M. Koca, Eur. Phys. J. \textbf{B48 }%
(2005) 583.

\bibitem{14} S. Cruz y Cruz, J. Negro and L.M. Nieto, Journal of Physics:
Conference Series \textbf{128, }(2008)\textbf{\ }012053.

\bibitem{15} S. Ghosh and S. K. Modak, Phys. Lett. A \textbf{373,} (2009)
1212.

\bibitem{16} B. Bagchi, S. Das, S. Ghosh and S. Poria, J. Phys. \textbf{A}:
Math. Theor. \textbf{46,} (2013) 032001.

\bibitem{17} S. H. Mazharimousevi, O. Mustafa, Phys. Scr. \textbf{87} (2013)
055008.

\bibitem{18} P. M. Mathews, M. Lakshmanan, Quart. Appl. Math. \textbf{32 }%
(1974)\textbf{\ }215.

\bibitem{19} A. Venkatesan, M. Lakshmanan, Phys. Rev. \textbf{E} \textbf{55}
(1997) 5134.

\bibitem{20} J. F. Cari\~{n}ena, M. F. Ra\~{n}ada, M. Santander, Regul.
Chaotic Dyn. \textbf{10} (2005) 423.

\bibitem{21} A. Bhuvaneswari, V. K. Chandrasekar, M. Santhilvelan, M.
Lakshmanan, J. Math. Phys. \textbf{53} (2012) 073504.

\bibitem{22} O. Mustafa, J. Phys. A: Math. Theor.\textbf{\ 46 }(2013) 368001.

\bibitem{23} A. K. Tiwari, S. N. Pandey, M. Santhilvelan, M. Lakshmanan, J.
Math. Phys. \textbf{54} (2013) 053506.

\bibitem{24} M. Lakshmanan, V. K. Chandrasekar, Eur. Phys J. ST \textbf{222}
(2013) 665.

\bibitem{25} Z. E. Musielak, J. Phys. \textbf{A}: Math. Theor. \textbf{41,}
(2008) 055205.

R. Bravo, M. S. Plyushchay, Phys. Rev. \textbf{D 93 }(2016) 105023.

\bibitem{26} C. Quesne, J. Math. Phys. \textbf{59} (2018) 042104; J. Math.
Phys. \textbf{56} (2015) 012903.

B. G. da Costa, E.P. Borges, J. Math. Phys. \textbf{59} (2018) 042101.

\bibitem{27} C. Quesne, V. M. Tkachuk, J. Phys. \textbf{A 37} (2004) 4267.

\bibitem{28} B. Bagchi, A. Banerjee, C. Quesne, V. M. Tkachuk, J. Phys. 
\textbf{A 38} (2005) 2929.

\bibitem{29} J. F. Cari\~{n}ena, M. F. Ra\~{n}ada, M. Santander, SIGMA 
\textbf{3} (2007) 030.

\bibitem{30} J. F. Cari\~{n}ena, M. F. Ra\~{n}ada, M. Santander, Ann. Phys. 
\textbf{322 }(2007) 434.

\bibitem{31} J. F. Cari\~{n}ena, M. F. Ra\~{n}ada, M. Santander, Rep. Math.
Phys. \textbf{54} (2004) 285.

\bibitem{32} C. Muriel, J. L. Romero, J. Phys. \textbf{A}: Math. Theor. 
\textbf{43} (2010) 434025.

\bibitem{33} R. G. Pradeep, V. K. Chandrasekar, M. Santhilvelan, M.
Lakshmanan, J. Math. Phys. \textbf{50} (2009) 052901.

\bibitem{34} N. Euler, M. Euler, J. Nonlinear Math. Phys. \textbf{11} (2004)
399.

\bibitem{35} K. S. Govinder, P. G. L. Leach, J. Math. Anal. Appl. \textbf{%
287 }(2003) 399.

\bibitem{36} N. Euler, T. Wolf, P. G. L. Leach, M. Euler, Acta Appl. Math. 
\textbf{76} (2003) 89.

\bibitem{37} J. F. Cari\~{n}ena, F. J. Herranz, M. F. Ra\~{n}ada, J. Math.
Phys. \textbf{58} (2017) 022701

\bibitem{38} M. Ranada, Phys. Lett. \textbf{A 379} (2015) 2267.

\bibitem{39} O. Mustafa, J. Phys. \textbf{A}; Math. Theor. \textbf{48}
(2015) 225206.

\bibitem{40} M. Ranada, M. A. Rodrigues, and M Santander, J. Math. Phys. 
\textbf{51}, (2010) 042901.

\bibitem{41} M. A. Rodrigues, p. Tempesta, and P. Winternitz, Phys. Rev. E 
\textbf{78,} (2008) 046608.

\bibitem{42} N. W. Evans and P. N. Verrier, J. Math. Phys. \textbf{49},
(2008) 092902.

\bibitem{43} M. Ranada, J. Math. Phys. \textbf{57}, (2016) 052703.

\bibitem{44} J. M. Jauch and E. L. Hill. Phys. Rev. \textbf{57}, (1940) 641.

\bibitem{45} M. Ranada, J. Math. Phys. \textbf{38}, (1997) 4165.

\bibitem{46} F. Tremblay, A. Turbiner, P. Winternitz, J. Phys. \textbf{A};
Math. Theor. \textbf{43} (2010) 015202.

\bibitem{47} W. Miller, S. Post, P. Winternitz, J. Phys. \textbf{A}; Math.
Theor. \textbf{46} (2013) 423001

\bibitem{48} R. C.-Stursberg, J. Math. Phys. \textbf{55} (2014) 042904.

\bibitem{49} O. Mustafa, S. H. Mazharimousavi, Int. J. Theor. Phys. \textbf{%
46} (2007) 1786.

\bibitem{50} O. Mustafa; arXiv:1208.2109:\textbf{\ }Comment on the
"Classical and quantum position-dependent mass harmonic oscillators" and
ordering-ambiguity resolution.

\bibitem{51} S. Gasirowicz, "\textit{Quantum Mechanics", 3rd Edition (}John
Wiley and Sons, Inc., New Jersey 2003).

\bibitem{52} O. Mustafa, S. H. Mazharimousavi, Int. J. Theor. Phys. \textbf{%
47} (2008) 1112.

\bibitem{53} O. Mustafa, S. H. Mazharimousavi, J. Phys. \textbf{A}: Math.
Theor. \textbf{40 }(2007) 863.

\bibitem{54} O. Mustafa, Int. J. Theor. Phys. \textbf{47} (2008) 1300.
\end{thebibliography}
\end{document}